\documentclass[sigconf, nonacm]{acmart}
\usepackage{graphicx}
\usepackage{lineno}
\usepackage{tabularx}
\usepackage{multirow}
\usepackage{multicol}
\usepackage{textcomp}
\usepackage{colortbl}
\usepackage{numprint}
\usepackage{textcomp}
\usepackage{array,tabularx}
\usepackage{booktabs}
\usepackage{enumitem}
\usepackage{textcomp}
\usepackage{comment}
\usepackage[utf8]{inputenc}
\usepackage{pdfpages}
\usepackage{array}
\usepackage{float}
\usepackage{subfig}
\usepackage{multirow}
\usepackage{graphics}
\usepackage{color}
\usepackage{algorithmic}
\usepackage{graphicx}
\usepackage{textcomp}
\usepackage{xcolor}
\usepackage{xspace}
\usepackage{geometry}
\usepackage{soul}
\usepackage{marginnote}
\usepackage[normalem]{ulem}
\usepackage{hyperref}
\usepackage{balance}
\usepackage{mdframed}
\usepackage{multicol}
\usepackage{multirow}
\usepackage{bookmark}

\newcommand{\ie}{\emph{i.e.,}\xspace}
\newcommand{\eg}{\emph{e.g.,}\xspace}

\newcommand{\etal}{\emph{et~al.}\xspace}

\newcommand{\CHANGED}[1]{#1}
\AtBeginDocument{%
  \providecommand\BibTeX{{%
    \normalfont B\kern-0.5em{\scshape i\kern-0.25em b}\kern-0.8em\TeX}}}

\copyrightyear{2022} 
\acmYear{2022} 
\setcopyright{acmlicensed}\acmConference[MSR '22]{19th International Conference on Mining Software Repositories}{May 23--24, 2022}{Pittsburgh, PA, USA}
\acmBooktitle{19th International Conference on Mining Software Repositories (MSR '22), May 23--24, 2022, Pittsburgh, PA, USA}
\acmPrice{15.00}
\acmDOI{10.1145/3524842.3527957}
\acmISBN{978-1-4503-9303-4/22/05}

\begin{document}

\title{How heated is it? Understanding GitHub locked issues}


\author{Isabella Ferreira}
\affiliation{%
  \institution{Polytechnique Montréal}
  \city{Montréal}
  \country{Canada}
}
\email{isabella.ferreira@polymtl.ca}

\author{Bram Adams}
\affiliation{%
  \institution{Queen's University}
  \city{Kingston}
  \country{Canada}
}
\email{bram.adams@queensu.ca}

\author{Jinghui Cheng}
\affiliation{%
  \institution{Polytechnique Montréal}
  \city{Montréal}
  \country{Canada}
}
\email{jinghui.cheng@polymtl.ca}

\renewcommand{\shortauthors}{Ferreira et al.}

\makeatletter
\newcommand\footnoteref[1]{\protected@xdef\@thefnmark{\ref{#1}}\@footnotemark}
\makeatother

\newcommand{\replicationPackage}{Replication package: \url{https://doi.org/10.6084/m9.figshare.18848765.v2}}

\begin{abstract}
Although issues of open source software are created to discuss and solve technical problems, conversations can become heated, with discussants getting angry and/or agitated for a variety of reasons, such as poor suggestions or  
violation of community conventions.
To prevent and mitigate discussions from getting heated, tools like GitHub have introduced the ability 
to lock issue discussions that violate the code of conduct or other community guidelines. Despite some early research on locked issues, there is a lack of understanding of how communities use this feature and of potential threats to validity 
for researchers relying on a dataset of locked issues as an oracle for heated discussions
. To address this gap, we (i) quantitatively analyzed 79 GitHub projects that have at least one issue locked as \textit{too heated}, and (ii) qualitatively analyzed all issues locked as \textit{too heated} of the 79 projects, a total of 205 issues comprising 5,511 comments. We found that projects have different behaviors when locking issues: while 54 locked less than 10\% of their closed issues, 14 projects locked more than 90\% of their closed issues. Additionally, \CHANGED{locked issues tend to have a similar number of comments, participants, and emoji reactions to non-locked issues}. For the 205 issues locked as \textit{too heated}, we found that one-third 
do not contain any uncivil discourse, and only 8.82\% of the analyzed comments are actually uncivil. Finally, we found that the locking justifications provided by maintainers do not always match the label used to lock the issue. Based on our results, 
we identified three pitfalls to avoid when using the GitHub locked issues data \CHANGED{and we provide recommendations for researchers and practitioners.}


\end{abstract}



\maketitle

\section{Introduction}



In open source software (OSS) development, 
community members use Issue Tracking Systems (ITSs) (\eg Jira, Bugzilla, and GitHub Issues) to discuss various topics related to their projects. 
Such ITSs provide a set of features that streamline communication and collaboration by promoting discussions around bug reports, requests of new features or enhancements, questions about the community and the project, and documentation feedback~\cite{heck2013analysis, arya2019analysis}.

Although issue reports play a crucial role in software development, maintenance, and evolution, issue discussions can get heated (or ``uncivil''), resulting in unnecessarily disrespectful conversations and personal attacks. 
This type of unhealthy, and sometimes disturbing or harmful behavior can be the result of a variety of reasons. For example, even though diversity has many benefits for open source communities~\cite{vasilescu2015gender, vasilescu2015perceptions}, the mix of cultures, personalities, and interests of open source contributors can cause a clash of personal values and opinions~\cite{cheng2019activity}. Furthermore, as a social-technical platform, ITSs sometimes host social context discussions, such as conversations about the \textit{black lives matter} and \textit{me too} movements, which can increase the chances of conflicts and arguments. Those discussions seek for a more anti-oppressive software terminology, such as renaming the branch \textit{master} to \textit{main}, \textit{whitelist/blacklist} to \textit{allowlist/blocklist} and gender-neutral pronouns. Finally, the increasing level of stress and burnout among OSS contributors can also cause unhealthy interactions~\cite{raman2020stress}. In fact, the amount of requests that OSS maintainers receive is overwhelming and the aggressive tone of some OSS interactions drains OSS developers~\cite{raman2020stress}.




Such heated interactions 
can have many negative consequences for OSS projects. Ferreira \etal~\cite{ferreira2021shut} have found that both maintainers and developers often discontinue further conversation and escalate the uncivil communication in code review discussions. Egelman \etal~\cite{egelman2020predicting} also found that interpersonal conflicts in code review can trigger negative emotions in developers. These communication styles might also hinder OSS communities' ability to attract, onboard, and retain contributors. 

To help OSS projects deal with some of the aforementioned challenges, in June 2014, GitHub released a feature that allows project owners 
to lock issues, pull requests, and commit conversations~\cite{holman_2014}, basically prohibiting further comments. The main goal of this feature is to smooth out \textit{too heated} conversations that violate the community's code of conduct~\cite{tourani2017code, li2021code} or GitHub's community guidelines~\cite{github_docs}. However, conversations can also be locked for other reasons, such as 
\textit{off-topic}, 
\textit{resolved}, or \textit{spam}. Contributors might also choose to lock issues without providing a reason.

\CHANGED{Since locked issues have been manually tagged by community experts rather than by researchers or classifiers, this data of locked issues provides a potentially valuable dataset for software engineering researchers aiming to understand how OSS communities handle possibly harmful conversations.}
A few very recent previous studies have used this dataset, in particular the subset of \textit{too heated} locked issues, as an oracle to detect toxicity in software engineering discussions~\cite{raman2020stress}, and to understand when, how, and why toxicity happens on GitHub locked issues~\cite{millerdid}. However, to the best of our knowledge, none of these studies have performed an in-depth investigation of the nature of GitHub locked issues in general and the validity of the \textit{too heated} locked issues in particular as a potential oracle.

Hence, in this paper, we adopt a mixed-methods approach and aim at assessing the characteristics of GitHub locked issues. 
First, we quantitatively analyzed 1,272,501 closed issue discussions of 79 open source projects hosted on GitHub that have at least one issue locked as \textit{too heated}. This analysis is aimed at identifying the overall characteristics of GitHub locked and non-locked issues.
Then, we qualitatively examined \textit{all} 205 issues locked as \textit{too heated} in the analyzed projects, and their 5,511 comments, to assess the extent to which the issue discussions locked as \textit{too heated} were, in fact, uncivil. For this
, we 
identified the tone-bearing discussion features (TBDFs) of issue discussions~\cite{ferreira2021shut}, \ie ``\textit{conversational characteristics demonstrated in a written sentence that convey a mood or style of expression}.'' We then identified heated discussions based on the presence of \textit{uncivil} TBDFs, \ie ``\textit{features of discussion that convey an unnecessarily disrespectful tone}.'' 
Additionally, we assessed the topics being discussed by \textit{too heated} locked issues and the justifications given by maintainers for locking such issues. 

In summary, we make the following contributions:
\begin{itemize}[noitemsep,topsep=0pt, leftmargin=20pt, itemindent=0pt]
    \item To the best of our knowledge, this is the first study shedding light on the usage patterns of the GitHub locking conversations feature; 
    \item We found that projects have different behaviors to lock issues, that the locking justifications given by maintainers do not always match the label on the GitHub platform, and that not all issues locked as \textit{too heated} are uncivil;
    \item We identified three pitfalls and provided a set of recommendations of what researchers should \textit{do} and \textit{not do} when using this dataset;
    \item \CHANGED{We provide three recommendations for practitioners and designers of ITSs;}
    \item We make a replication package\footnote{\label{replicationpackage}\replicationPackage} available 
      that contains (i) the codebooks used in the qualitative coding, (ii) the manually tagged dataset of issues locked as \textit{too heated}, containing sentences coded with TBDFs, the topics of discussion, and the justifications given by maintainers, \CHANGED{and (iii) the scripts to analyze the data.}
\end{itemize}

\section{Related work: Unhealthy Interactions in OSS}

Our work is related to a few very recent studies~\cite{ferreira2021shut, millerdid, cheriyan2021towards, sarker2020benchmark, raman2020stress} that investigate unhealthy interactions, mostly in the form of \textit{incivility} and \textit{toxicity}, that happen in OSS contexts. Ferreira \etal~\cite{ferreira2021shut} investigated incivility (\ie ``\textit{heated discussions that involve unnecessary disrespectful comments and personal attacks}'') in code review discussions of rejected patches of the Linux Kernel Mailing List. They found that incivility was present in 66\% of the non-technical code review emails of rejected patches; they also identified that frustration, name calling, and impatience were the most common types of incivility. Miller \etal~\cite{millerdid} investigated toxicity, which is defined as ``\textit{rude, disrespectful, or unreasonable language that is likely to make someone leave a discussion}'' in 100 open source issue discussions (including 20 GitHub issues locked as \textit{too heated}). The authors found that entitlement, insults, and arrogance were the most common types of toxicity in issue discussions. 

Because the nature of discussions is different, previous work found different causes of unhealthy interactions in code review and issue discussions. In code review discussions, incivility was found to be mainly caused by the violation of community conventions, inappropriate solutions proposed by the developers, and characteristics of the maintainers' feedback~\cite{ferreira2021shut}. However, in issue discussions, users have been found to write toxic comments when having problems using the software, holding different points of view about technical topics, or being in a disagreement about politics and ideology (\eg OSS philosophy)~\cite{millerdid}.

These studies also identified patterns of reactions from OSS communities to unhealthy interactions. In code reviews, maintainers often discontinue further conversation or escalate uncivil communication~\cite{ferreira2021shut}. In issue discussions, although maintainers and other members may try to engage in a constructive conversation after toxicity happens, the discussion might still escalate to more toxicity~\cite{millerdid}. However, these discussions seemed to be localized; only in a few cases, the author who posted toxic comments would open another toxic issue~\cite{millerdid}.
Additionally, when maintainers invoke the code of conduct~\cite{tourani2017code}, the author of the toxic comments usually did not comment anymore. These previous results indicate that locking issues might be effective to stop toxicity. Our study is built upon these recent works to focus on understanding (i) how GitHub locked issues are used in practice and (ii) to what extent GitHub issues locked as \textit{too heated} 
are indeed uncivil.

Researchers have also worked on building classifiers to detect unhealthy interactions. Recently, Cheriyan \etal~\cite{cheriyan2021towards} have proposed a machine learning-based method to detect and classify offensive comments into swearing and profanity. The classifier is trained with a dataset including conversations in GitHub Issues, Gitter, Slack, and Stack Overflow. Raman \etal~\cite{raman2020stress} created a classifier to automatically detect toxicity on GitHub. The authors manually validated GitHub issues locked as \textit{too heated} to train the classifier; their classifier obtained a satisfiable precision but low recall.

Finally, Sarker \etal~\cite{sarker2020benchmark} analyzed how different toxicity detectors perform on software engineering data, finding that none of the analyzed detectors achieved a good performance. However, the design of such detectors often lacks an in-depth understanding of the characteristics of the dataset. We contribute to filling this gap by investigating the reliability and pitfalls of using the GitHub locked issues data, in particular \textit{too heated} locked issues, to train machine learning detectors of incivility. 


\section{Study Design}


\subsection{Study goal and research questions}

The general goal of this study is to understand the nature of GitHub locked issues and to identify the common pitfalls that could pose a threat to validity when using a (sub)set of GitHub locked issues as an oracle for uncivil communication. Our specific goals are to (i) quantitatively characterize GitHub locked issues in comparison to non-locked issues
and (ii) qualitatively assess the actual discussion tones in issues locked as \textit{too heated}. Based on these goals, we constructed four main research questions to guide our study, which we present below along with their motivations.\\

\textbf{\textit{RQ1. What are the characteristics of GitHub locked issues?}}
To the best of our knowledge, only the study conducted by Miller \etal~\cite{millerdid} has studied GitHub locked issues, in particular focusing on \textit{when}, \textit{how}, and \textit{why} toxicity happens on a sample of 20 locked issues. In our quantitative study, we aim to conduct a broader analysis of locked issues to identify their overall characteristics.
It is essential to gain this understanding to make informed decisions about mining this kind of data and to understand how different open source projects use this feature. To this end, we assess how often projects lock issues, as well as how different locked issues are in comparison to issues that are not locked, in terms of the number of comments, the number of people participating in the discussion, and the number of emoji reactions.\\ 


\textbf{\textit{RQ2. What are the justifications given in the comments by project maintainers when locking issues as \textit{too heated}?}}
Project maintainers can choose predefined reasons (\eg \textit{too heated} or \textit{spam}) to lock an issue on GitHub. However, these predefined reasons are abstract and sometimes difficult to interpret. Maintainers often need to communicate the specific justifications of locking a too-heated issue with the community in the issue comments in order to explain their actions, educate the community members, and/or maintain their authority. In this RQ, we aim at understanding how maintainers communicate these justifications.\\


\textbf{\textit{RQ3. What are the topics being discussed in issues locked as \textit{too heated}?}}
Earlier work by Ferreira \etal~\cite{ferreira2021shut} did not find any correlation between the topics of code review discussions and the presence of incivility. However, this finding might not necessarily apply to issue discussions because code review and issues discussions have different focuses (the former on the solution space while the latter on the problem space) and participant dynamics (there is a smaller power distance between maintainers and other discussants in issue discussions than in code reviews). Thus, in this RQ, we aim at examining the topics of discussion in issues locked as \textit{too heated} in order to analyze the presence of potentially provocative topics.\\

\textbf{\textit{RQ4. To what extent are issues locked as \textit{too heated} uncivil?}}
In this RQ, we aim at investigating the extent to which issues locked as \textit{too heated} do, in fact, involve heated interactions. We used the characterization of incivility 
of Ferreira \etal~\cite{ferreira2021shut} to identify the 
  uncivil tones in issue discussions as a concrete measure of heated discussions. Our analyses were split into four sub-RQs.

First, we aim at answering \textit{\textbf{RQ4.1. What are the features of discussion in issues locked as \textit{too heated}?}} This is done by identifying the tone-bearing discussion features (TBDFs) of the sentences of these issues
. TBDFs capture the ``\textit{conversational characteristics demonstrated in a written sentence that convey a mood or style of expression}''~\cite{ferreira2021shut}. 
As an example of a TBDF, the following sentence shows that a speaker is frustrated: ``\textit{I'm fed up the whole framework is filled with this crap. Why you even do double way binding if works half the needed cases? Don't even make a framework at this point.}''  (\texttt{project angular/angular}), demonstrating therefore a mood or style of expression. 
Because the original TBDF framework was created through analysis of code reviews, we adapted this framework to the context of issue discussions, then used this adapted framework to identify the TBDFs in each sentence of the issue discussions in our sample.

Then, 
we answer \textit{\textbf{RQ4.2. How uncivil are issues locked as \textit{too heated}?}}. For this, we use the notion of \textit{uncivil TBDFs}, which are ``\textit{features of discussion that convey an unnecessarily disrespectful tone}''~\cite{ferreira2021shut}. We consider an issue or comment as \textit{technical} if none of its sentences demonstrate any TBDF in 
RQ4.1, \textit{uncivil} if at least one sentence demonstrates an uncivil TBDF, and \textit{civil} if at least one sentence demonstrates a TBDF 
but none of these TBDFs are uncivil
. Conceptually, uncivil issues and comments correspond to inappropriate and unhealthy discussions.

Finally, we use the aforementioned categorization of issues and comments to assess correlations between (1) the TBDFs demonstrated in an issue discussion and (2) justifications given by project contributors for locking an issue and the topics of the \textit{too heated} issues. Thus we ask: \textit{\textbf{RQ4.3. How are the observed discussion feature types distributed across the justifications given by project contributors when locking \textit{too heated} issues?}} and \textit{\textbf{RQ4.4. How are the observed discussion feature types distributed across the different discussion topics in \textit{too heated} issues?}} 

\subsection{Data selection}

Our research questions require open source projects with a sufficient number and variety of locked issues. Since no pre-built dataset exists, we used a two-pronged approach. First of all, we selected 
projects from 32 different GitHub collections\footnote{\url{https://github.com/collections}}, which are curated lists of diverse and influential GitHub projects. We selected projects that (i) have more than 1,000 issues, (ii) have code commits later than November 2020 (six months before our data collection), and (iii) have at least one issue locked as \textit{too heated} until 2021-06-03. This resulted in 29 GitHub projects.

Second, we also added 
projects open-sourced by three large, well-known software companies (Apple, Google, and Microsoft), since we hypothesized that such projects might encounter more polarized discussions. For this, we mined the companies' corresponding GitHub organization, but also added open-sourced projects not hosted within the organization (such as Bazel and Kubernetes). This resulted in 3,931 projects,
including 86 projects from Apple, 1,240 projects from Google, and 2,605 projects from Microsoft. 
Filtering out projects without issues locked as \textit{too heated} until 2021-06-03 resulted in 50 projects, \ie 1 
project from Apple, 23 
from Google, and 26 
from Microsoft.

Then, we collected all closed issues from the resulting sample of 79 projects until 2021-06-03 using the GitHub REST API\footnote{\url{https://docs.github.com/en/rest/reference/issues}}. We chose to analyze only closed issues in order to observe complete issue discussions. For each closed issue, we recorded whether the issue has been locked~\cite{github_docs_lock}, the issue comments, the emoji reactions to each comment, all the events related to the issue (\eg when the issue was locked), the dates that the issue was opened, closed, and locked (if so), and the contributors who performed these actions. 





\subsection{Quantitative analysis on locked issues}
To answer RQ1, we considered two independent groups: locked and non-locked issues, as well as three dependent variables. \CHANGED{We discuss the hypothesis related to each dependent variable below.}

\CHANGED{
\textbf{Number of comments:} Ferreira \etal found that review discussions with arguments tend to be longer and have more uncivil than civil emails~\cite{ferreira2021shut}. We thus hypothesize that locked issue discussions have more comments than non-locked discussions (H1).

\textbf{Number of participants:} Previous research found that discussions involving people with different backgrounds and cultures are more likely to be uncivil~\cite{salminen2020topic}. Since OSS projects are highly diverse, we hypothesize that locked issues have more participants than non-locked issues (H2).

\textbf{Number of emoji reactions:} Previous work has shown that emoji reactions reduce unnecessary commenting in pull request discussions, leading to fewer conflicts~\cite{son2021more}. We thus hypothesize that locked issues have fewer emoji reactions than non-locked issues (H3).
}

\CHANGED{We aggregate the dependent variables per issue for each independent group. Following the central limit theorem~\cite{islam2018sample} and because of the large number of issues in our dataset (about 1.3 million), we used unpaired t-tests~\cite{whitley2002statistics} to determine if there is a significant difference in the means of each dependent variable between the two independent groups. Then, we computed the effect size between the means of the two groups for each dependent variable using Cohen's d~\cite{rice2005comparing}.}



\subsection{Qualitative analysis on locked issues}
To answer RQ2, RQ3, and RQ4, we conducted a qualitative analysis~\cite{strauss1987qualitative, strauss1990open} on the GitHub issues locked as \textit{too heated}. 

\subsubsection{Identifying the justifications to lock GitHub issues.}
\label{section:identifying_justifications}

In RQ2, we aim at investigating the justifications given by community members themselves when locking issues as \textit{too heated}. 
We approached the coding through the following steps. First, we read the title and the first comment (description) of the issue to understand what the issue is about. Then, we read the last comment of the issue. If the last comment mentions that the issue is being locked and it is clear \textit{why} it is the end of the discussion, we then coded for the justification given by the community member. If the last comment does not mention that the issue is being locked or it is not clear \textit{why}, we then searched for other comments justifying the reason for locking the issue, as follows.
\begin{enumerate}[noitemsep, leftmargin=*, itemindent=0pt]
    \item We searched for the keywords ``locking'', ``locked'', ``closing'', ``heated''. If we found a comment mentioning one of these keywords, we read the comment and check if it mentions the justification for locking the issue. If not, we executed step (2).
    \item We read the entire issue thread looking for a justification to lock the issue. If we did not find any justification, we then coded the justification as ``\textit{no reason mentioned}''. We coded the respective justification, otherwise.
\end{enumerate}

The first author started with an inductive coding~\cite{thomas2003general} on all issues locked as \textit{too heated}, a total of 205 issues. During the coding process, we added the identified justifications in a codebook~\cite{saldana_coding} with the name of the code, a definition, and one or more examples. The codebook containing all manually identified justifications was improved during discussions with two other authors. Then, we conducted axial coding and grouped the identified justifications into themes~\cite{vollstedt2019introduction}.

To guarantee that the codebook can be replicated, the second author deductively coded 20\%~\cite{distaso2012multi, foundjem2019release}
of the issues (41 issues). Afterwards, the first and third authors discussed the disagreements, improving our coding schema.
We computed Cohen's Kappa to evaluate the inter-rater reliability of our coding schema~\cite{mcdonald2019reliability}. The average Kappa score between the two raters across all the identified justifications is 0.85, ranging from 0.64 to 1.00, demonstrating an almost perfect agreement~\cite{viera2005understanding}. 
The complete codebook can be found in our replication package\footnote{\label{replicationpackage}\replicationPackage}.

\subsubsection{Identifying the topic of the discussion} For RQ3, we first inductively coded the issue title and the content being discussed at the issue level. 
We then did axial coding to group our codes into categories. A codebook was created with the axial codes and their definitions. Two authors discussed and iteratively improved the codebook, which can be found in our replication package\footnoteref{replicationpackage}. 

\subsubsection{Identifying tone-bearing discussion features (TBDFs).} 
We identify the discourse characteristics of issues locked as \textit{too heated} (RQ4) through tone-bearing discussion features (TBDFs)~\cite{ferreira2021shut}. 
To initiate the coding, the first author directly used the framework and the codebook of 16 TBDFs proposed by Ferreira~\etal~\cite{ferreira2021shut}. 
To adapt the coding schema in the issue discussion context, we added new codes or adjusted the coding criteria of certain codes when appropriate. The coding was conducted at the sentence level of each issue comment written in English and visible on GitHub (a total of 5,511 comments from the 205 issues locked as \textit{too heated}). 
We also took into consideration the context of the previous comments on the same issue to identify the TBDF of a particular sentence.

The updated codebook was iteratively discussed and improved with all authors. In the end, we added four TBDFs and adjusted the coding criteria of eight TBDFs in the original codebook. The third author then deductively coded 20\%~\cite{distaso2012multi, foundjem2019release} of the comments coded with at least one TBDF by the first author (145 out of 718 comments). We measured the inter-rater reliability, and the average Cohen's Kappa score between the two raters and across all the identified TBDFs was 0.65, ranging from 0.43 to 0.91, showing substantial agreement between the two raters~\cite{viera2005understanding}. 
The final codebook can be found in our replication package\footnoteref{replicationpackage}.

\section{Results}


\subsection{RQ1. Characteristics of GitHub locked issues}



\begin{figure}[t]
\centering
\includegraphics[clip, width=0.75\linewidth]{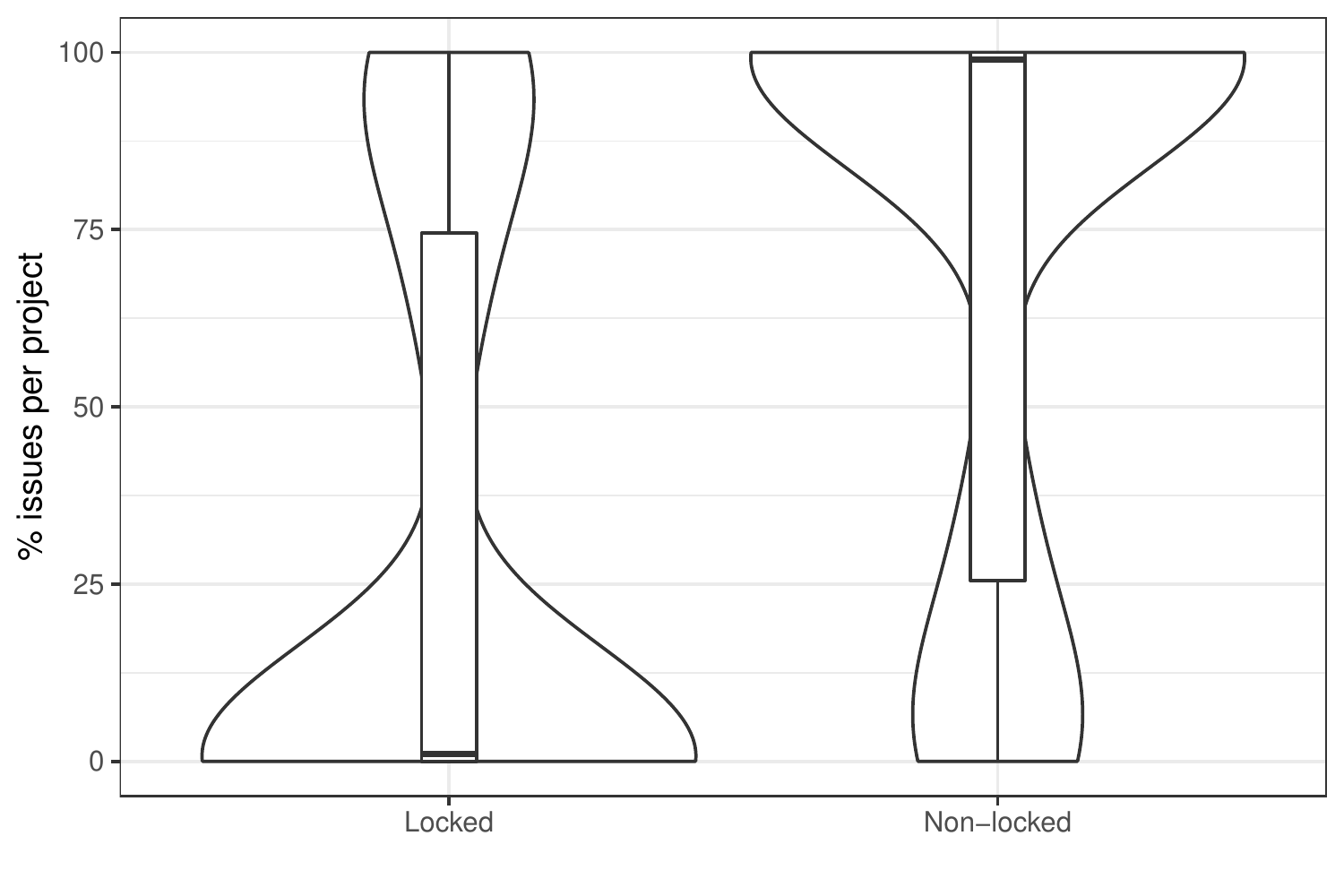}
\caption{Distribution of percentages of (non-)locked issues per project.}
\label{fig:distribution_locked_issues_1}
\end{figure}
\begin{figure}[t]
\centering
\includegraphics[clip, width=0.75\linewidth]{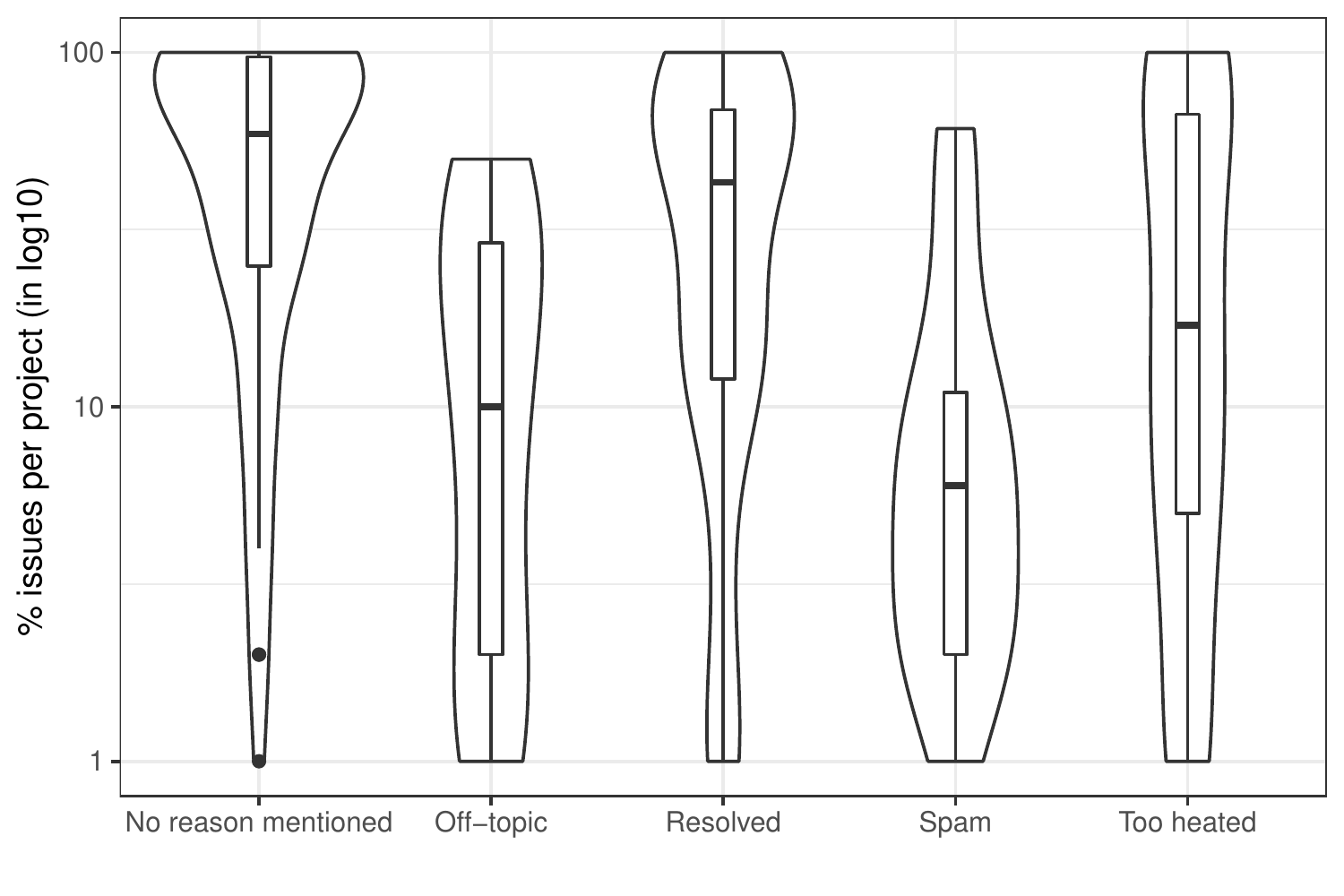}
\caption{Distribution of percentages of (non-)locked issues per project according to locking reasons labeled on GitHub.}
\label{fig:distribution_locked_issues_2}
\end{figure}

Among the 1,272,501 closed issues of the 79 analyzed projects, \textbf{we identified three types of projects: 14 projects  (17.72\%) locked more than 90\% of their closed issues, 54 projects (68.35\%) locked less than 10\% of their closed issues; the remaining 11 projects (13.92\%) locked between 54\% and 88\% of their closed issues with an average of 73\% of locked issues.} Figures~\ref{fig:distribution_locked_issues_1} and \ref{fig:distribution_locked_issues_2} present the distribution of (non-)locked issues per project as well as per locking reason mentioned on the GitHub platform, respectively.



Furthermore, \textbf{we found that 313,731 (61.61\%) locked issues have been automatically locked by a bot (\eg due to inactivity), 195,409 (38.38\%) by an organization, and 52 (0.01\%) by a user}. Interestingly, 16 projects (20.25\%) had most of their issues locked with no reason mentioned and 9 projects (11.4\%) had most of their issues locked as resolved. These 25 projects, specifically, have 313,494 (62.13\% of their locked issues) issues locked by a bot.

 
\CHANGED{Concerning the length of locked and non-locked issues,
an unpaired t-test revealed that non-locked issues had a statistically significantly larger number of comments ($mean=5.72$, $SD=10.54$) than locked issues ($mean=5.05$, $SD=8.23$), $t=-40.587, p < 0.001$. However, the difference between the means of these two variables is negligible (Cohen’s $d = 0.07$), \textbf{rejecting H1}. Similarly, we found that non-locked issues involved a statistically significantly larger number of participants ($mean=2.92$, $SD=2.48$) than locked issues ($mean=2.86$, $SD=3.21$), $t=-11.559, p < 0.001$. However, Cohen's $d=0.02$ indicated that this difference is negligible, \textbf{rejecting H2}. Finally, we found that non-locked issues involved a statistically significantly larger number of reactions ($mean=0.13$, $SD=0.85$) than locked issues ($mean=0.07$, $SD=0.60$), $t=-40.631, p < 0.001$, but this difference is again negligible (Cohen's $d=0.07$), \textbf{rejecting H3}.}\\

\begin{mdframed}[backgroundcolor=black!10]
    \textbf{Summary RQ1}: 
    We found three types of projects in terms of issue locking behaviors: (1) 14 projects locked more than 90\% of their closed issues, (2) 54 locked less than 10\%, and (3) the remaining 11 locked between 54\% and 88\% of their closed issues.
    Furthermore, 31.65\% of the projects have the majority of issues locked by a bot. Finally, \CHANGED{locked issues tended to have a similar number of comments, participants, and emoji reactions to non-locked issues.}
    
\end{mdframed}
\subsection{RQ2. Justifications for locking GitHub issues as \textit{too heated}}
\label{sec:justification-rq2}




In 70 issues (34.15\% of the 205 \textit{too heated} locked issues), the justification for locking the issue was not explicitly mentioned by the project contributors and not clear from the discussion. 
In the remaining 135 issues, \textbf{we found ten categories of justifications that project contributors gave when locking the issues as \textit{too heated}.}

\textbf{\textit{Inappropriate/unhealthy interaction}} was the justification for locking 52 issues (25.37\%). Contributors locked these issues by calling out behaviors violating the code of conduct, asking other contributors to keep the discourse civil, or mentioning that the kind of behavior will not be tolerated by the community. Furthermore, project contributors locked issues because the conversation was starting to become uncivil, and in a few cases, the offensive comments were even hidden
. As an example: ``\textit{This is not the place to post this kind of message @[username], I'm closing the topic. Please follow the contributing guidelines if you want to post anything constructive.}'' (\texttt{project angular/angular}).

\textbf{\textit{Off-topic.}} Project contributors explicitly mentioned that they were locking 23 issues (11.22\%) because the discussion was getting off-topic. This includes cases where the issues were not actionable or unrelated to the project goals, and/or people were discussing the implications related to other issues but not the issue being discussed. \textit{E.g.,} ``\textit{Thanks for creating this issue. We think this issue is unactionable or unrelated to the goals of this project. Please follow our issue reporting guidelines.}'' (\texttt{project microsoft/vscode}).

\textbf{\textit{Issue will not be addressed.}} The justification for locking 16 issues (7.80\%) was that the team decided not to address the issue. This can be due to various reasons such as different motivations between the team and the users, disagreement about licensing, geopolitical or racial concerns, or project contribution process problems. \textit{E.g.,} ``\textit{We discussed this issue on the SIG-arch call of 20200130, and have unanimously agreed that we will keep the current naming for the aforementioned reasons.}'' (\texttt{project kubernetes/kubernetes}).

\textbf{\textit{Issue/PR status.}} The reason for locking 11 issues (5.37\%) was due to the issue or pull request status: duplicated, merged, not mergeable, inactive, stale, fixed, abandoned, etc. In most cases in this category, project contributors locked the issue instead of closing it. \textit{E.g.,} ``\textit{This PR is being closed because golang.org/cl/281212 has been abandoned.}'' (\texttt{project golang/go}).

\textbf{\textit{Wrong communication channel}} was the justification for locking 7 issues (3.41\%). In this case, the contributor mentioned that the problem should be discussed in another channel, such as the mailing list or the IRC channel. \textit{E.g.,} ``\textit{The community site is where we are moving conversations about problems happening on travis-ci.com or travis-ci.org. Thanks in advance for posting your questions over there.}'' (\texttt{project travis-ci/travis-ci}).

\textbf{\textit{Issue cannot be addressed.}} Project contributors locked 7 issues (3.41\%) because it was not feasible to address the issue under discussion. The underlying reason could be that the discussion did not provide a reasonable way to address the problem, the issue is related to another project or a dependency, or the project does not have enough resources (such as personnel or infrastructure) to address the issue. In most cases, issues in this category were locked as \textit{too heated} instead of closed. \textit{E.g.,} ``\textit{I'm going to close this issue (and edit the OP for fast reference) since the issue is deeper in the OS and the Flutter framework can't resolve it.}'' (\texttt{project flutter/flutter}).

\textbf{\textit{Work prioritization.}} Project contributors locked 6 issues (2.93\%) to save time answering discussions and requests, or to focus on other work aspects. \textit{E.g.,} ``\textit{Closing this conversation to prevent more ``ETA requested'' responses (which actually take time from feature work)}'' (\texttt{project firebase/FirebaseUI-Android}).

\textbf{\textit{Not following community rules.}} Project contributors locked 6 issues (2.93\%) because contributors were not following the community rules. For example, contributors used $+1$ comments instead of the emoji reaction +1, discussed too many problems in one issue, or continued the discussion on issues with a similar topic. \textit{E.g.,} ``\textit{I've locked this to contributors for now. Adding +1 comments is too noisy. For future reference, add a reaction to the issue body, and don't comment.}'' (\texttt{project ansible/ansible}).

\textbf{\textit{Address the issue in the future.}} Project contributors justified the reason for locking 5 issues (2.44\%) being that the problem will be addressed in the future. That is, the contributor mentioned that the bug will be triaged with other bugs, the issue needs further investigation, or the bug will be addressed in the future. \textit{E.g.,} ``\textit{As always, thanks for reporting this. We'll definitely be triaging this with the rest of our bug fixes. In the meantime, however, please keep the discourse civil.}'' (\texttt{project microsoft/terminal}).

\textbf{\textit{Communication problems.}} There were 2 issues (0.97\%) locked because the discussion was not being productive or people could not reach a consensus. \textit{E.g.,} ``\textit{I'm going to close this thread, as the conversation isn't really productive after [link to a comment], I'm afraid.}'' (\texttt{project flutter/flutter}).\\


\begin{mdframed}[backgroundcolor=black!10]
    \textbf{Summary RQ2:} We identified ten justifications that project contributors gave when locking issues as \textit{too heated}. In the majority of the issues (74.63\%), the justifications were \textit{not} related to the conversation being uncivil.
\end{mdframed}
\subsection{RQ3. Topics of discussions in issues locked as \textit{too heated}}
In 12 issues (5.85\%) that were locked as being \textit{too heated}, we were not able to identify the discussion topic from the issue title or the comments. In the remaining 193 issues, \textbf{we found 13 topics}:

\textbf{\textit{Source code problems}} was the topic of 78 issues (38.05\%). Examples of such problems include deprecated functionality, encoding problems, code warning, and installation problems.

\textbf{\textit{User interface}} is the topic of 35 issue discussions (17.07\%). Contributors were concerned with the interface colors, the user interface crashing or not responsive, or the need for changing icons.

\textbf{\textit{Renaming.}} There were 15 issues (7.32\%) discussing about renaming the software, the API, or certain terminologies due to racial concerns, such as renaming \textit{master} to \textit{main} and \textit{whitelist} to \textit{allowlist}.

\textbf{\textit{New feature.}} There were 15 issues (7.32\%) locked as \textit{too heated} that were discussing the implementation of a new feature. In some cases, the project was asking the community about feature ideas for the next releases. In other cases, people were requesting to add social features (such as Instagram and Twitter) to the terminal and to consider specific syntax in the source code.

\textbf{\textit{Community/project management.}} We found 12 issues (5.85\%) discussing topics related to the community and project management. More specifically, contributors were asking the project's owner to give more privileges to other people to review and merge code, criticizing censorship when removing comments and locking issues, asking questions about coding programs such as \textit{freeCodeCamp} and \textit{Google Summer of Code}, discussing about project financing, making announcements about the end of the project, etc.

\textbf{\textit{Lack of accordance.}} Contributors expressed their opinion to not work for a specific company or their opinion about a tool or programming language in 9 issues (4.39\%).

\textbf{\textit{Data collection/protection.}} We found 7 issues (3.41\%) discussing data collection or data protection. More specifically, contributors were concerned about the security, privacy, and ethical issues related to data collection, storage, and sharing. 

\textbf{\textit{Documentation.}} The topic of 6 issues (2.93\%) was related to documentation, such as errors, missing information, or inappropriate content (\eg political banners) in the documentation. 

\textbf{\textit{Performance}} is the topic of 5 issues (2.44\%). Contributors were discussing the performance problems between two releases, the application or the download of the application was very slow, and interactivity with the webpage was very slow.

\textbf{\textit{Error handling.}} 4 issues (1.95\%) discussed compilation errors, tool errors, or errors after a version update.

\textbf{\textit{Translation.}} We found 3 issues (1.46\%) discussing a problem with the listing of languages in the software (\eg listing Chinese (China), and Chinese (Taiwan) separately), or regarding languages from the Google Translator (\eg Scottish Gaelic).

\textbf{\textit{Versioning.}} There were 2 issues (0.98\%) about requests to change the version number of the tool or complaints about the tool not following semantic versioning.

\textbf{\textit{License.}} There were 2 issues (0.98\%) about making changes in the license file.\\





\begin{mdframed}[backgroundcolor=black!10]
    \textbf{Summary RQ3}: We found 13 topics being discussed in issues locked as \textit{too heated}, with \textit{source code problems}, \textit{user interface}, and \textit{renaming} being the most frequent topics.
\end{mdframed}
\subsection{RQ4. Incivility in issues locked as \textit{too heated}}

In this RQ, we aim at assessing to what extent issues locked as \textit{too heated} feature uncivil discourse. We present the results of each sub-RQ below.


\subsubsection{RQ4.1. What are the features of discussion in issues locked as \textit{too heated}?}
\label{section:tbdfs}

\textbf{We identified 20 tone-bearing discussion features (TBDFs) in issue discussions locked as \textit{too heated}, 
four of which have not been found by previous work.} 
\textbf{In total, 1,212 distinct sentences were coded with a TBDF} (a sentence can be coded with more than one TBDF). We present below the description, an example, and the frequency of the four TBDFs uniquely identified in the analyzed issue discussions. For the 16 TBDFs identified by Ferreira \etal~\cite{ferreira2021shut}, we present the frequency and the conditions that were added to code such TBDFs when different from previous work. For replication purposes, the description and the example of TBDFs not described here can be found in our replication package\footnote{\label{replicationpackage}\replicationPackage}.\\

\noindent
\textit{Positive features.} \textbf{Surprisingly, 151 sentences (12.46\%) in issue discussions locked as \textit{too heated} actually expressed a positive tone}. \textbf{\textit{Considerateness}} is the most frequent positive feature ($N=61$), followed by \textbf{\textit{Appreciation and excitement}} ($N=58$), and \textbf{\textit{Humility}} ($N=32$). Contributors also expressed \textit{Appreciation and excitement} towards the project in issue discussions; \eg ``\textit{You have a super reliable, and extendable set of tools. wcf provides the best tools for building enterprise applications. Without it, you simply end up rebuilding it.}'' (\texttt{project dotnet/wcf}).\\




\noindent
\textit{Neutral features} \textbf{appear in 115 sentences (9.49\%) of issue discussions locked as \textit{too heated}.} In the context of issue discussions, we have identified \textbf{\textit{Expectation}} ($N=44$) and \textbf{\textit{Confusion}} ($N=12$) as two new neutral TBDFs. Additionally, \textbf{\textit{Sincere apologies}} ($N=31$), \textbf{\textit{Friendly joke}} ($N=25$), and \textbf{\textit{Hope to get feedback}} ($N=3$) also appear in issue discussions, similar to code review discussions~\cite{ferreira2021shut}.

\textbf{\textit{Expectation}} is a new TBDF we identified in the issue discussions and is the most frequent neutral feature in our dataset ($N=44$). This TBDF is expressed when the speaker expects to add a feature in the future, to fix a specific problem, or that the feature should do something specific. It is also expressed when the speaker is expecting someone to resolve a problem or that the community will work on a particular problem. \textit{E.g.,} ``\textit{As a consumer of your product, I expect it to work as advertised.}'' (\texttt{project jekyll/jekyll}).



We also identified \textbf{\textit{Confusion}}, which is expressed when the speaker is unable to think clearly or understand something ($N=12$). \textit{E.g.,} ``\textit{I am confused because I add this `mixins` to as own which should not affect any updates from Bootstrap.}'' (\texttt{project twbs/bootstrap}).\\


\noindent
\textit{Negative features.} \textbf{There were 196 distinct sentences (16.17\%) demonstrating negative features.} While indicating a negative mood, these TBDFs do not involve a disrespectful tone. In our dataset, we found \textbf{\textit{Commanding}} ($N=61$), \textbf{\textit{Sadness}} ($N=40$), and \textbf{\textit{Oppression}} ($N=9$), which were already identified by previous work~\cite{ferreira2021shut}. However, different from code reviews, in issue discussions contributors also expressed \textit{Sadness} when the community is going to lose something or someone (\eg ``\textit{It would be a great loss for the .NET community if you'd stop contributing.}'' \texttt{project dotnet\_runtime}), and \textit{Oppression} when a person of power reinforces their standpoints (\eg ``\textit{bro I am an Open Source author and maintainer so don't try lecturing me about being against "off-putting towards the open-source community.}'' \texttt{project dotnet/maui}). Additionally, we identified two new negative TBDFs: \textbf{\textit{Dissatisfaction}} ($N=75$) and \textbf{\textit{Criticizing oppression}} ($N=17$)

\textbf{\textit{Dissatisfaction}} appears when a simple change requires a lot of discussions and the change is not accepted, when someone wants to stop contributing because things never get resolved, or the community does not acknowledge the problem or is not willing to fix the problem. \textit{E.g.,} ``\textit{At this point I am discouraged to report more because nothing ever seems to get fixed (usually because ``it's complicated").}'' (\texttt{project dotnet/roslyn}). Additionally, contributors might express dissatisfaction with the framework, tool, or process. 



\textbf{\textit{Criticizing oppression}} happens when someone of a lesser power (\eg developer) does not accept what someone (usually a person of a higher power) says or how the person behaves. \textit{E.g.,} ``\textit{Your heavy-handed and dismissive approach to moderation diminishes the project and the whole community.}'' (\texttt{project nodejs/node}).\\


\textit{Uncivil features.}
Uncivil features are those that convey an unnecessarily disrespectful tone~\cite{ferreira2021shut}. \textbf{We identified 790 sentences (65.18\%) featuring at least one uncivil TBDF.} \textbf{\textit{Annoyance and Bitter frustration}} is the most common uncivil feature ($N=288$). Issue discussions also demonstrate \textbf{\textit{Name calling}} ($N=222$), \textbf{\textit{Mocking}} ($N=194$), \textbf{\textit{Irony}} ($N=64$), \textbf{\textit{Impatience}} ($N=55$), \textbf{\textit{Vulgarity}} ($N=51$), and \textbf{\textit{Threat}} ($N=18$). Although contributors express these uncivil TBDFs in both code review discussions~\cite{ferreira2021shut} and issue discussions, we found that several TBDFs had new interpretations in the context of issue discussions, which we describe below.

Contributors tend to express \textbf{\textit{Annoyance and Bitter frustration}} in issue discussions when they use capital letters to emphasize something in a frustrating way, when someone is using abusive language to express their opinion, when injustice makes the other person feel unable to defend herself/himself, and when the speaker is strongly irritated by something impossible to do in the speaker's opinion. Contributors might also mention that they are ``fed up'', ``pissed off'', ``sick'', and ``tired'' of something. \textit{E.g.,} ``\textit{I'm not just an angry fool, a lot of people are fed up with it and it actually is a half-baked crippled tool.}'' (\texttt{project angular/angular}).

\textbf{\textit{Name calling}} was expressed in issue discussions by mentioning ``you'', the name or identification of someone on GitHub (usually expressed by \texttt{@username}), or the name of a company in a sentence that has a negative connotation. \textit{E.g.,} ``\textit{Obviously you didn't take my hint on length to heart but don't come to conclusions about my character because you don't know me at all.}'' (\texttt{project flutter/flutter}).

Contributors expressed \textbf{\textit{Mocking}} in issue discussions by making fun of the community rules or mimicking the way someone speaks. \textit{E.g.,} ``\textit{People can express their opinions but the mods are just gonna lock it away ``cause y'all can't behave'' and ``the discussion is getting out of hand and is not productive anymore''.}'' (\texttt{project angular/angular}).

Contributors expressed \textbf{\textit{Impatience}} when other community members asked them to work on a bug even if they do not have enough resources, when they are unhappy with the situation that exists for a long time, and when someone comments before reading and understanding the message. \textit{E.g.,} ``\textit{As I have stated multiple times you can definitely work with Atom while offline (or with network requests blocked).}'' (\texttt{project atom/atom}).

\textbf{\textit{Threat}} is demonstrated in issue discussions when someone mentions that a person will be punished if they do not follow the code of conduct, when contributors threaten to stop using a product, or when someone is challenging someone else. \textit{E.g.,} ``\textit{I have the urge to drop Microsoft products and suggest to the company I work for that we do the same wherever we can.}'' (\texttt{project dotnet/roslyn}).\\

\begin{mdframed}[backgroundcolor=black!10]
    \textbf{Summary RQ4.1:} We identified 20 TBDFs in issues locked as \textit{too heated}, four of which have never been found by previous work. \textit{Annoyance and bitter frustration}, \textit{name calling}, and \textit{mocking} are the most common TBDFs in the analyzed issues. 
\end{mdframed}


\subsubsection{RQ4.2. How uncivil are issues locked as \textit{too heated}?}
Building on the sentence-level TBDF coding of RQ4.1, we then consider the overall issue or comment as \textbf{civil} if it contains sentences coded with \textit{positive}, \textit{neutral}, and/or \textit{negative} features. An issue/comment is considered \textbf{uncivil} if it contains sentences coded with \textit{at least} one \textit{uncivil} TBDF. Finally, an issue/comment is considered \textbf{technical} if none of its sentences are coded with a TBDF, \ie the issue discussion is focused only on technical aspects.

\begin{figure}[t]
\centering
\includegraphics[clip, width=0.70\linewidth]{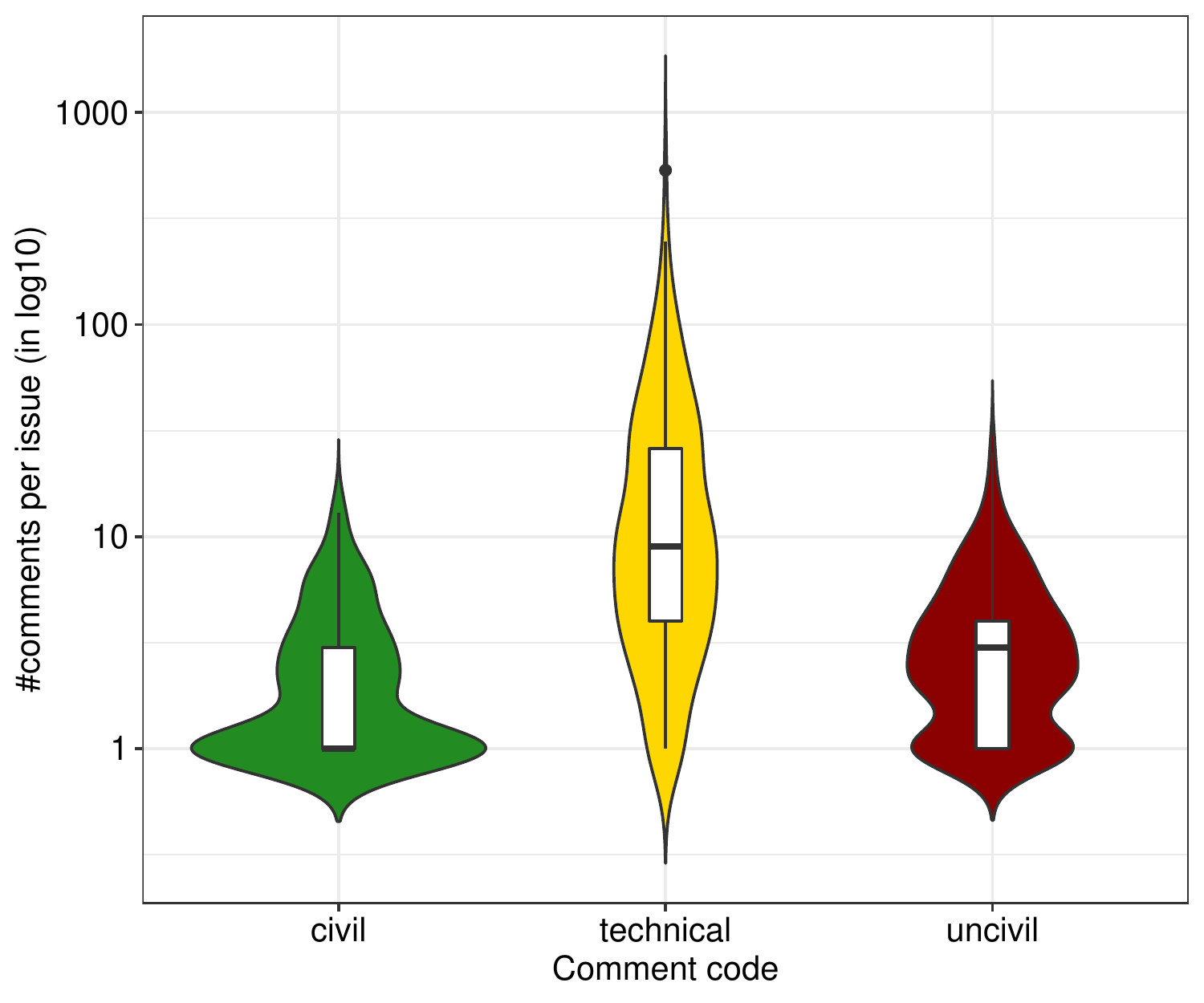}
\caption{Distribution of the frequency of the three types of comments across issues.}
\label{fig:boxplot_email_code}
\end{figure}

\textbf{Discussions locked as \textit{too heated} can still have civil comments or even involve civil or technical comments only.} 
From the 205 issues locked as \textit{too heated}, 138 (67.32\%) of them are uncivil, 45 (21.95\%) are technical, and 22 (10.73\%) issues are civil. From the 5,511 comments part of the 205 issues, 4,793 (86.97\%) of them are technical, 486 (8.82\%) are uncivil, and 232 (4.21\%) are civil. We also observe that the median numbers of technical, uncivil, and civil comments in issues locked as \textit{too heated} are 9, 1, and 0, respectively. Figure~\ref{fig:boxplot_email_code} presents the distribution of the number of comments per issue of the three types of comments. 

Inspired by the coding framework of Miller \etal~\cite{millerdid}, we investigate where the uncivil comments are positioned in uncivil issues. Particularly, we considered three locations: (1) in the issue description, (2) in the first comment, and (3) in later comments (\ie emerged from the discussion). For each one of the 138 issues that included at least one uncivil comment, the combination of the above three locations resulted in seven conditions of where the uncivil comments were positioned: (i) only in the issue description, (ii) only in the first comment, (iii) only in the issue description and the first comment, (iv) in the issue description and it emerged from the discussion, (v) in the first comment and it emerged from the discussion, (vi) in the issue description, first comment, and it emerged from the discussion, or (vii) only emerged from the discussion.

\begin{figure}[t]
\centering
\includegraphics[clip, width=0.9\linewidth]{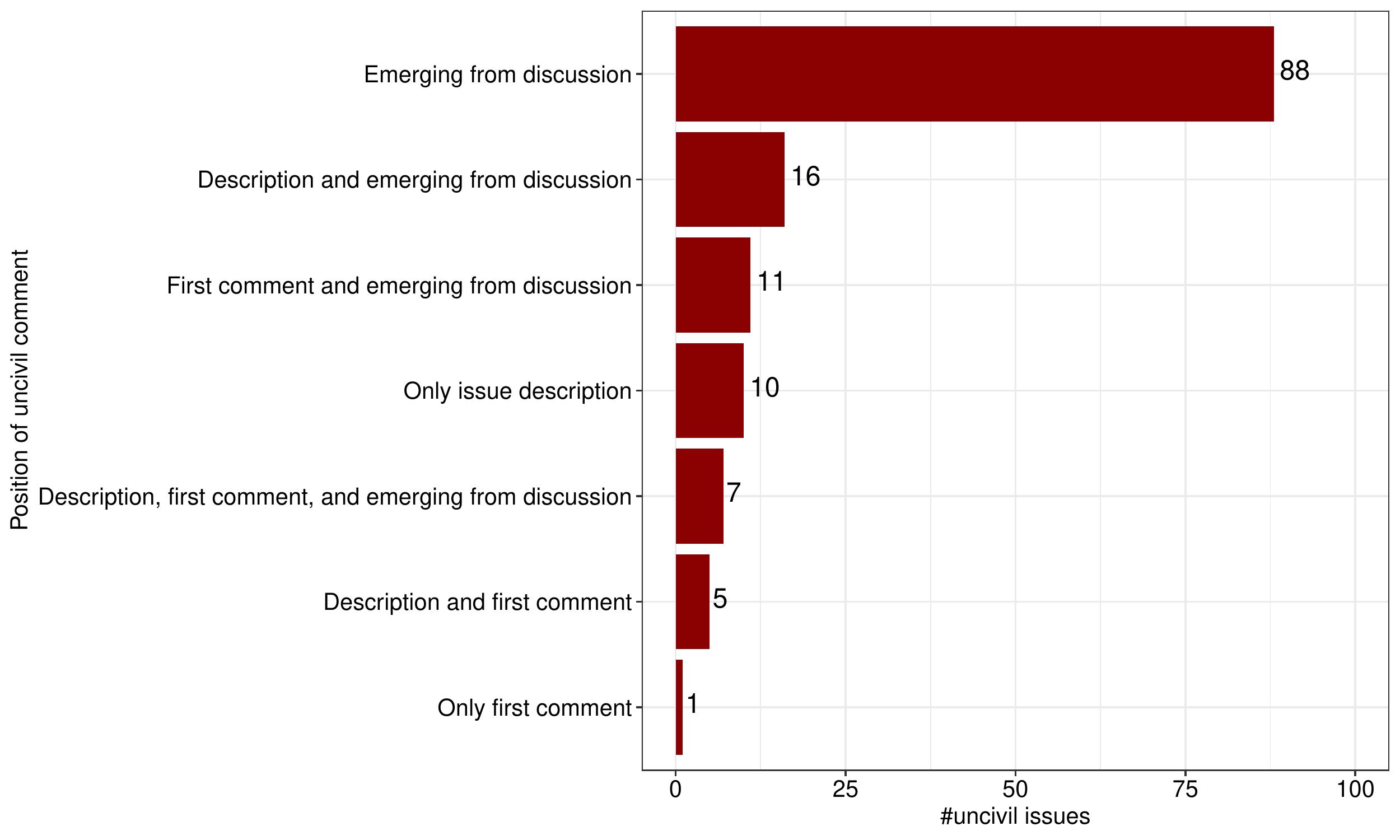}
\caption{Position of uncivil comments in uncivil issues.}
\label{fig:barplot_position_uncivil_comments}
\end{figure}

As shown in Figure~\ref{fig:barplot_position_uncivil_comments}, \textbf{uncivil comments emerged from the discussion in 88 issues (63.77\%)}, incivility was present on the issue description and it emerged from the discussion in 16 issues (11.59\%), and it was present on the first comment and emerged from the discussion in 10 issues (7.25\%).\\

\begin{mdframed}[backgroundcolor=black!10]
    \textbf{Summary RQ4.2:} Contrary to expectations, 32.68\% of issues locked as \textit{too heated} are either technical or civil, and only 8.82\% of the comments in issues locked as \textit{too heated} are uncivil. Additionally, uncivil comments emerge from the actual discussion in 63.77\% of the uncivil issues.
\end{mdframed}






\subsubsection{RQ4.3. How are the observed TBDF types distributed across the locking justifications?}



\begin{figure}[t]
\centering
\includegraphics[clip, width=0.85\linewidth]{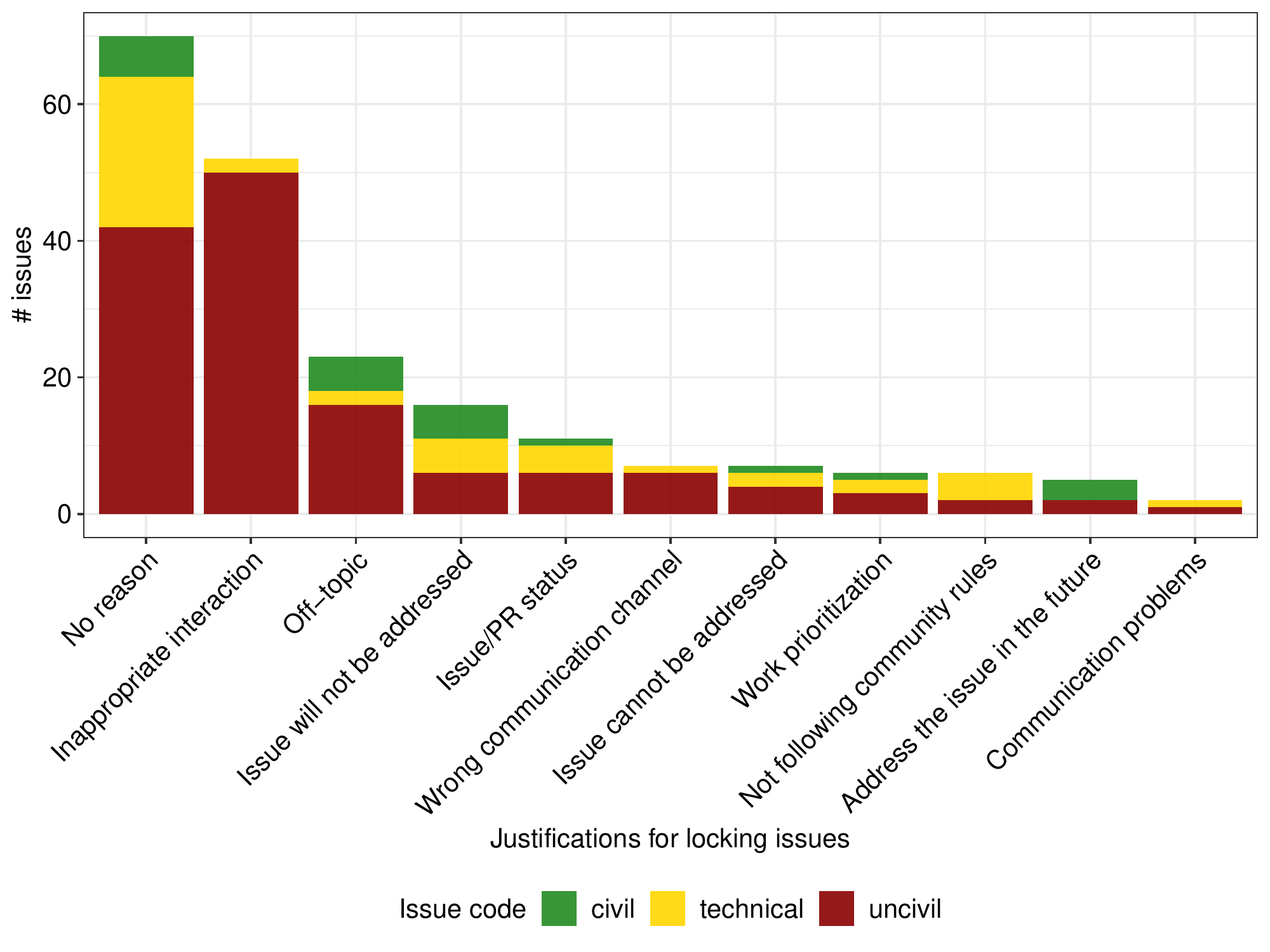}
\caption{Number of issues per justifications given by maintainers when locking issues as \textit{too heated}.}
\label{fig:locking_reason_issue_level}
\end{figure}

As observed in Figure~\ref{fig:locking_reason_issue_level}, although project contributors did not mention a justification for locking 70 issues, 60\% of those issues (42 issues) were \textit{uncivil}; the remaining 28 issues (40\%) were either \textit{technical} (22 issues) or civil (6 issues). \textbf{This result shows that, in this case, the lack of justification given by the project contributors is not a reliable reason to filter out such data, since they still contain a high number of uncivil issues.}

As expected, 96.15\% of the issues locked with a justification of \textit{inappropriate/unhealthy interaction} included one or more \textit{uncivil} comments, while none were 
\textit{civil}, and two issues (3.85\%) were \textit{technical}. These findings might be due to the fact that project contributors can delete or hide heated comments, which we did not consider in our analysis. When analyzing the \textit{off-topic} justification for \textit{too heated} locked issues, we found that 69.57\% (16) of those issues were uncivil, 21.74\% (5) were civil, and 8.70\% (2) were technical. In this case, the latter seven issues should have been locked by the project contributor using the \textit{off-topic} option, instead of \textit{too heated}.

\begin{figure}[t]
\centering
\includegraphics[clip, width=0.9\linewidth]{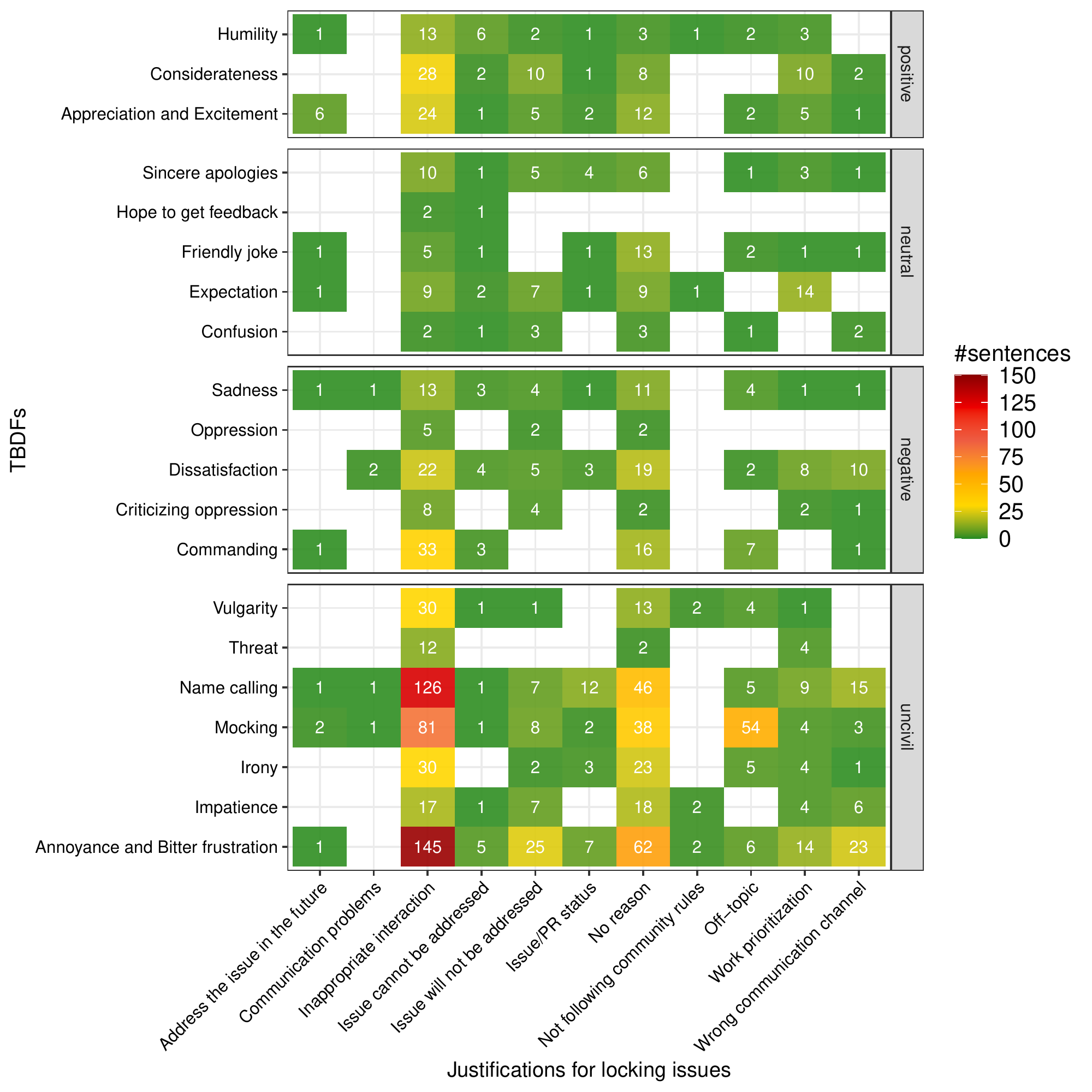}
\caption{Justifications given by maintainers when locking issues as \textit{too heated} per TBDF.}
\label{fig:heatmap_justifications_tbdf}
\end{figure}

\textbf{Issues locked due to \textit{inappropriate/unhealthy interaction} often demonstrate \textit{annoyance and bitter frustration}, \textit{mocking}, and \textit{name calling}.} Figure~\ref{fig:heatmap_justifications_tbdf} presents the frequency of TBDFs per identified justification given by maintainers when locking issues as \textit{too heated}. Issues locked with \textit{no reason mentioned} often demonstrate \textit{annoyance and bitter frustration}, \textit{mocking}, and \textit{name calling}. Interestingly, \textit{too heated} issues locked because the conversation was \textit{off-topic} often demonstrate \textit{mocking}. This is because most of the issues in this category are from the project \texttt{microsoft/vscode} and are related to the Santagate event~\cite{santagate}. To celebrate the holiday season, the VS Code team added a Santa hat to the settings gear. A user wrote an issue complaining that the Santa hat was very offensive to him. After that, the VS Code team changed the icon to a snowflake. However, many users got frustrated with that change and started writing a lot of issues and comments filled with \textit{mocking}.\\

\begin{mdframed}[backgroundcolor=black!10]
    \textbf{Summary RQ4.3:} Although all identified justifications for \textit{too heated} issues contain some proportion of issues with incivility, contributors only called out for uncivil communication in 25.37\% of all issues, representing 37.68\% of the uncivil issues. Furthermore, 60\% of the issues in which project contributors did not mention a justification for locking the issue are actually uncivil.
\end{mdframed}



\subsubsection{RQ4.4. How are the observed TBDF types distributed across the discussion topics?}
\begin{figure}[t]
\centering
\includegraphics[clip, width=0.85\linewidth]{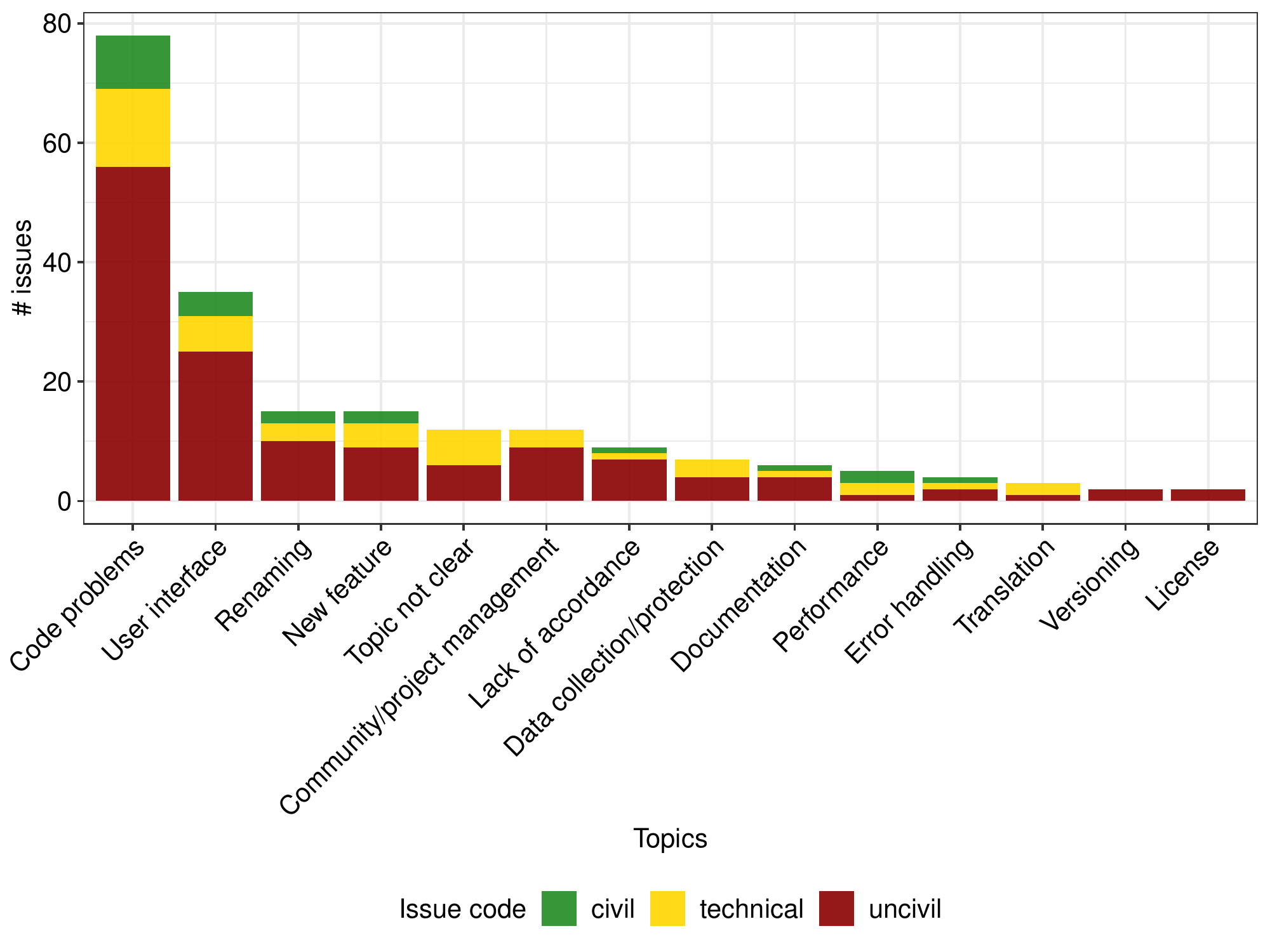}
\caption{Number of issues per topics of issues locked as \textit{too heated}.}
\label{fig:topic_issue_level}
\end{figure}

Figure~\ref{fig:topic_issue_level} shows the number of civil, technical, and uncivil issues per topic of discussion. Although 56 issues (71.79\%) discussing \textit{source code problems} are uncivil, 22 issues (28.25\%) are either technical (13 issues) or civil (9 issues). A similar pattern is observed for the topics \textit{user interface} and \textit{renaming}.

\textbf{Interestingly, none of the issues discussing \textit{community/project management}, \textit{data collection/protection}, \textit{translation}, and \textit{topic not clear} exhibit civility}, \ie issues discussing the aforementioned topics are either \textit{uncivil} or \textit{technical}. \textbf{All issues discussing \textit{versioning} and \textit{license} are uncivil}, and \textbf{issues discussing \textit{performance} and \textit{translation} are more technical and civil than uncivil} (6 technical/civil issues (75\%) against 2 uncivil issues (25\%)).

\begin{figure}[t]
\centering
\includegraphics[clip, width=0.9\linewidth]{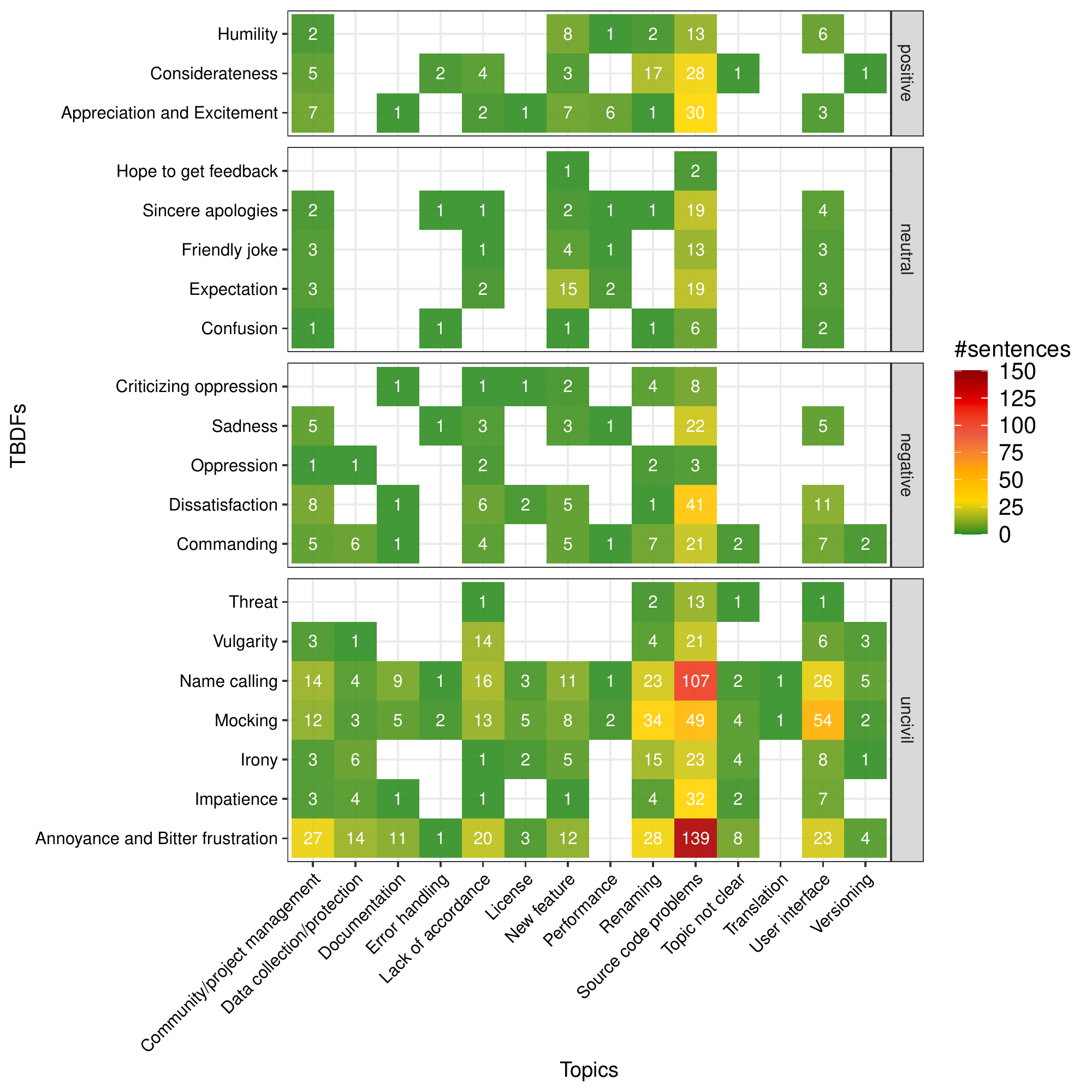}
\caption{Topics of issues locked as \textit{too heated} per TBDF.}
\label{fig:heatmap_topics_tbdf}
\end{figure}

Figure~\ref{fig:heatmap_topics_tbdf} presents the frequency of topics per TBDF. \textbf{We found that issues discussing \textit{source code problems} often demonstrate \textit{annoyance and bitter frustration} and \textit{name calling}.} 

\begin{mdframed}[backgroundcolor=black!10]
    \textbf{Summary RQ4.4:} All \textit{too heated} issues discussing \textit{versioning} and \textit{license} were uncivil, and issues discussing \textit{performance} and \textit{translation} tended to be more technical or civil. 
\end{mdframed}

\section{Discussion and Recommendations}

\CHANGED{We present below a discussion about (i) how projects are using the GitHub locking conversations feature, (ii) how incivility is expressed in issues locked as \textit{too heated}, and (iii) recommendations for researchers and practitioners.}






\subsection{How are projects using the GitHub locking conversations feature?} 

We found that projects have different behaviors when locking their issues. In fact, 14 projects locked more than 90\% of their closed issues, while 54 locked less than 10\% of their closed issues
. 
Furthermore, the overall percentage of locked issues (40.02\%) was surprising to us; this might be due to the fact that 61.61\% of the locked issues were automatically locked by a bot (\eg due to inactivity). We also found that project maintainers give a variety of justifications to the community when locking issues as \textit{too heated}, while most of the justifications are not related to the issue being uncivil. 

In fact, it seems that issues locked because the conversation was \textit{off-topic} should have been locked as ``off-topic'' instead, and issues locked with other justifications (such as \textit{Issue/PR status}, \textit{Issue will not be addressed}, or \textit{Issue cannot be addressed}) should have been closed normally instead of locked. Furthermore, we found that issues locked as \textit{too heated} often focused on discussing topics related to \textit{source code problems}, \textit{renaming}, and \textit{user interface}. For discussions about source code problems, more specifically, if participants abstain from demonstrating \textit{annoyance and bitter frustration}, \textit{name calling}, and \textit{mocking}, the conversation would be more civil. Some topics seem to not trigger incivility though, such as \textit{performance} and \textit{translation}.

\noindent
\subsection{How is incivility expressed in issues locked as \textit{too heated}?} 
Similar to code review discussions on LKML~\cite{ferreira2021shut}, we found that issue discussions often demonstrate \textit{annoyance and bitter frustration} and \textit{name calling}. Furthermore, we found that issue discussions feature \textit{expectation}, \textit{confusion}, \textit{dissatisfaction}, and \textit{criticizing oppression}, which have not been found in code review discussions of the LKML~\cite{ferreira2021shut}. That said, \textit{confusion} was investigated in code review discussions of Android, Facebook, and Twitter projects~\cite{ebert2019confusion}. These differences are most likely a result of the distinct ways in which issues and code reviews are discussed (\ie issues focus frequently on the problem space while code reviews focus on the solution space) as well as the communication style of different OSS communities.

Finally, we found that in most (63.77\%) of the issues that included an uncivil comment, those comments emerged during the discussion, instead of appearing at the beginning of the issue. This might happen due to a variety of reasons. In code review discussions, uncivil comments emerge from the discussion when maintainers do not immediately answer or developers do not solve the problem with the provided information~\cite{millerdid}, or the maintainer's feedback is inadequate, there is a violation of community conventions, or poor code quality~\cite{ferreira2021shut}. However, the causes of incivility in locked issues are still unknown and we encourage further studies to investigate this direction.



\subsection{\CHANGED{Recommendations}}

\CHANGED{Based on our results, we provide a set of recommendations for both researchers and practitioners.}

\subsubsection {\CHANGED{For researchers}}

We have identified three potential pitfalls for researchers that use GitHub locked issues data ``out of the box'', and we provide recommendations to mitigate these problems.\\

\begin{mdframed}[backgroundcolor=black!10]
    \textbf{Pitfall 1}: Bots automatically lock issues. 
\end{mdframed}
\vspace{6pt}

About one-third (31.65\%) of the projects in our sample had the majority (62.13\%) of closed issues locked by a bot (\eg the Lock Threads bot\footnote{\url{https://github.com/dessant/lock-threads}} that locks issues due to inactivity), instead of a maintainer. 
Thus, using the GitHub locked issues data as it is might be misleading, since issues locked as resolved do not necessarily mean that the issue is indeed resolved. In fact, the GitHub guidelines~\cite{github_docs_lock} recommend that issues should be locked when the conversation is not constructive or violates the project's code of conduct or GitHub's community guidelines, while the bots often lock issues for reasons other than these.

\textbf{Recommendations:} 
\textbf{Don't} use all locked issues assuming that the project is following the GitHub's guidelines. \textbf{Do} use the GitHub events~\cite{github_docs_event} to verify if the issues have been locked by a bot. If so, the real reason why the issue is locked should be examined (manually).  \\

\begin{mdframed}[backgroundcolor=black!10]
    \textbf{Pitfall 2}: Issues locked as \textit{too heated} may not contain uncivil expressions.
\end{mdframed}
\vspace{6pt}

According to our analysis of incivility, 32.68\% of the issues locked as \textit{too heated} are either technical or civil; \ie they do not contain any uncivil comment. We hypothesize that maintainers locked such issues as \textit{too heated} to prevent the discussion from becoming heated. As a result, researchers should not assume that uncivil expressions would necessarily appear in issues locked as \textit{too heated}. In fact, we found that the majority of the comments (91.18\%) in those issues are not heated/uncivil at all. Thus, blindly using this dataset (\eg directly feeding it to a machine learning model to detect inappropriate behavior) might lead to unreliable results.

\textbf{Recommendations:}
\textbf{Don't} blindly use the issues locked as \textit{too heated} assuming that they all include uncivil or \textit{too heated} discussions.
\textbf{Do} manually inspect the data to identify heated/uncivil comments and construct manually annotated datasets for automated techniques. To reduce this effort, benchmarks could be built and reused.\\


\begin{mdframed}[backgroundcolor=black!10]
    \textbf{Pitfall 3}: The justification given by maintainers may not match the label they used to lock the issues or reflect the true locking reason.
\end{mdframed}
\vspace{6pt}

For all the issues locked with a \textit{too heated} label, maintainers only explicitly justified 25.37\% of them with a comment on inappropriate/unhealthy interaction. For the other 74.63\%, the justifications given were, among others, \textit{off-topic}, \textit{issue will not be addressed}, or \textit{issue/PR status}. Furthermore, among the 70 issues to which the maintainers did not explicitly provide a justification for locking, we found that the majority (60\%) in fact include an uncivil comment.
There are different explanations of why maintainers provide justifications other than the reason they used to lock the issue. They might have locked the issue before it gets \textit{too heated} as a preventive measure, provided a nebulous justification to avoid confrontation, or simply chosen the wrong reason for locking. Researchers should consider these factors when using the labels or the explicitly given justifications.

\textbf{Recommendations:}
\textbf{Don't} blindly accept either the justification label and/or text as the true oracle for why an issue is locked.  
\textbf{Do} scrutinize factors such as the discussion topic, the context of the discussion, and the presence/absence of unhealthy interaction.

\subsubsection{\CHANGED{For practitioners}}

\CHANGED{We suggest the following recommendations for practitioners and designers of issue tracking systems (ITSs).}\\

\begin{mdframed}[backgroundcolor=black!10]
\CHANGED{\textbf{Recommendation 1}: Projects should have clear and explicit guidelines for maintainers to lock issues according to each locking reason.}
\end{mdframed}
\vspace{6pt}

\CHANGED{We identified three types of projects that locked issues, \ie projects that locked (i) more than 90\% of their closed issues, (ii) less than 10\% of their closed issues, and (iii) between 54\% and 88\% of their closed issues. Hence, having explicit guidelines would not only guarantee consistency amongst the maintainers of a given OSS project but would also ensure transparency to the entire community.\\}

\begin{mdframed}[backgroundcolor=black!10]
\CHANGED{\textbf{Recommendation 2}: Projects should not abuse the locking issue feature (\eg locking instead of closing issues).}
\end{mdframed}
\vspace{6pt}

\CHANGED{According to the GitHub guidelines~\cite{github_docs}, conversations should only be locked when they are not constructive. However, we found that 17.72\% of projects locked more than 90\% of their closed issues. This could have an adverse effect on the project since OSS contributors might assume at first sight that the community is uncivil.\\}

\begin{mdframed}[backgroundcolor=black!10]
\CHANGED{\textbf{Recommendation 3}: ITSs should provide features that (i) allow projects to add custom locking reasons, (ii) allow maintainers to select more than one locking reason (\eg spam and too heated), and (iii) encourage maintainers to add a justification of why the issue is being locked.}
\end{mdframed}
\vspace{6pt}

\CHANGED{We found that maintainers give different justifications when locking issues and that 
such justifications do not match the label on the GitHub platform in 74.63\% of the issues locked as \textit{too heated}. Furthermore, maintainers did not mention a justification for locking 70 issues as \textit{too heated}, out of which we could not observe signs of incivility in 28 issues. Finally, some \textit{too heated} issues are also \textit{spam}, such as issues related to the Santagate event. New features in ITSs are necessary to mitigate the aforementioned problems.
}


\section{Threats to Validity}


\textbf{Construct validity.}
We used incivility, particularly Ferreira \etal's framework of TBDFs~\cite{ferreira2021shut}, as a proxy to identify and measure heated discussions and expressions. While this is the most appropriate framework we found adaptable to our context, incivility defined in this framework may not completely overlap with the concept of heated discussions. 

\textbf{Internal validity.}
First, our qualitative coding can lead to inconsistencies due to its subjectiveness. To minimize this threat, our codebooks were interactively improved with all three authors. Additionally, we validated our codings with a second rater, reaching an almost perfect agreement for the coding of justifications given by maintainers and a substantial agreement for the TBDFs. Second, we only coded for comments that are visible on the GitHub platform and were not able to analyze hidden or deleted comments. So it is possible that issues coded as technical or civil actually included hidden or deleted comments that were uncivil. 
Third, we only qualitatively analyzed issues 
locked as \textit{too heated}, but since adding a reason to lock issues is optional on the GitHub platform, we might have missed heated issues that were not explicitly labeled by a maintainer with a reason or that were labeled with another reason. 

\textbf{External validity.}
Although the projects were carefully selected and filtered based on a set of criteria, our results are based on a sample of 79 projects that have at least one issue locked as \textit{too heated} in the analyzed period. Hence, our results cannot be generalized to projects that do not have any issues locked as \textit{too heated}. To minimize this threat, we compared locked and non-locked issues in our quantitative analyses and we qualitatively assessed all comments in \textit{all} issues locked as \textit{too heated}. Furthermore, the selected projects might not be the ones with the largest number of locked issues. Although this is a threat to the study validity, we could still find interesting insights in the analyzed sample.


\section{Conclusion}
In this paper, we focus on an empirical study aimed at understanding the characteristics of the locked issues on GitHub, particularly those locked as \textit{too heated}. In our sample of 79 projects, we found that projects have different behaviors when it comes to locking issues, and that 
\CHANGED{locked issues tend to have a similar number of comments, participants, and emoji reactions to non-locked issues.} Through an analysis of the 205 issues locked as \textit{too heated} in our dataset, we found that the locking justifications provided by the maintainers in the comments do not always match the label used to lock the issue. The topics covered by those issues are also diverse. Leveraging a framework capturing uncivil behavior in software engineering discussions, we identified that about one-third of the issues locked as \textit{too heated} do not contain any uncivil comments. Our analysis also revealed patterns in how the civil and uncivil discussion features are distributed across the explicit justifications and the discussion topics. Together, our results provide a detailed overview of how GitHub's locking conversations feature is used in practice, and we suggest concrete pitfalls and recommendations for software engineering researchers and practitioners using this information.

\section*{Acknowledgements}
The authors thank the Natural Sciences and Engineering Research Council of Canada for funding this research through the Discovery Grants Program [RGPIN-2018-04470].

\bibliographystyle{ACM-Reference-Format}
\bibliography{bibliography.bib}

\end{document}